\documentclass[twocolumn,showpacs,preprintnumbers,amsmath,amssymb]{revtex4}
\usepackage{graphicx}
\usepackage{dcolumn}
\usepackage{bm}

\newcommand{\be}{\begin{equation}}
\newcommand{\ee}{\end{equation}}
\newcommand{\bea}{\begin{eqnarray}}
\newcommand{\eea}{\end{eqnarray}}

\newcommand{\ti}{\times}
\newcommand{\half}{\frac{1}{2}}

\newcommand{\mc}{\mathcal}
\newcommand{\ph}{\phantom}
\newcommand{\nn}{\nonumber}

\begin{document}
\preprint{DAMTP-2012-59}
\title{Dark Radiation in LARGE Volume Models}
\author{Michele Cicoli$^{*,**}$, Joseph P. Conlon$^{\dagger}$, Fernando Quevedo$^{*, \ddagger}$}
\affiliation{$\ph{g}$ \\
$^{*}$ ICTP, Strada Costiera 11, Trieste 34014, Italy \\
$^{**}$INFN, Sezione di Trieste, Italy\\
$^{\dagger}$ Rudolf Peierls Centre for Theoretical Physics, 1 Keble Road, Oxford, OX1 3NP, UK \\
$^{\ddagger}$ DAMTP, Centre for Mathematical Sciences, Wilberforce Road, Cambridge, CB3 0WA}
\begin{abstract}
We consider reheating driven by volume modulus decays in the LARGE Volume Scenario. Such reheating always generates
non-zero dark radiation through the decays to the axion partner, while the only competitive visible sector decays are Higgs pairs via the
Giudice-Masiero term. In the framework of sequestered models where the cosmological moduli problem is absent, the simplest model
with a shift-symmetric Higgs sector generates $1.56\leq \Delta N_{eff}\leq 1.74$. For more general cases, the known experimental bounds on
$\Delta N_{eff}$ strongly constrain the parameters and matter content of the models.
\end{abstract}
\maketitle

\section{Introduction}

The cosmological Standard Model (SM) starts with a period of inflation. During this period, the energy of the universe is dominated by the vacuum energy of a slowly rolling scalar field. At some point inflation ends and, irrespective of the overall particle spectrum or number of hidden sectors,
the energy has to be transferred predominantly into thermal relativistic SM degrees of freedom via a process of reheating.

Constraints on this are measured via $N_{eff}$, the effective number of neutrino species.
$N_{eff}$ is measured both at BBN and CMB times,
and in practice measures the fraction of the total energy density that lies in the thermal photon plasma.
At CMB temperatures $N_{eff}$ is determined in terms of the total energy by
\be
\rho_{total} = \rho_{\gamma} \left( 1 + \frac{7}{8}\left( \frac{4}{11} \right)^{4/3} N_{eff} \right).
\ee
In the SM $N_{eff,BBN} = 3$ and $N_{eff,CMB} = 3.04$, due to partial reheating of the neutrinos from $e^{+} e^{-}$ annihilation.
The presence of additional dark radiation, decoupled from the SM and relativistic at both BBN and CMB temperatures,
leads to $\Delta N_{eff} \equiv N_{eff} - N_{eff,SM} > 0$.

Observations show a mild but consistent preference for $\Delta N_{eff} > 0$.
At CMB times WMAP, ACT and SPT report $N_{eff} = 4.34^{+0.86}_{-0.88},$  $4.56 \pm 0.75,  3.86 \pm 0.42$ respectively \cite{wmapetc}.
At BBN times an excess has also been reported but the evidence depends on the relic Helium abundance \cite{manganoetc}. Based only
on D/H, \cite{NollettHolder} reports $N_{eff} = 3.9 \pm 0.44$. A recent general overview is \cite{steigman}.

As the inflationary universe is vacuum energy dominated, dark radiation must arise during or after reheating.
In the context of string models of the early universe, there are two main challenges in
understanding reheating. The first, the cosmological moduli problem (CMP) \cite{cmp} is to understand how re-\emph{heat}-ing can occur at all, and the second is
to ensure that it is primarily the SM that is reheated.

We recall first the CMP \cite{cmp}. String theory
contains many moduli associated to the complicated Calabi-Yau geometry. Moduli are typically Planck-coupled scalars which are expected to obtain vevs during inflation, leading to post-inflationary production of moduli through the vacuum misalignment mechanism.
 Moduli oscillate coherently as matter, redshift slowly and come to dominate the energy density of the universe. As Planck-coupled fields, their characteristic decay rate
is $\Gamma \sim \frac{1}{16 \pi} \frac{m_{\phi}^3}{M_P^2}$. A reheating temperature $T \gtrsim \mc{O}(1)\, \hbox{MeV}$, necessary for BBN, then requires $m_{\phi} \gtrsim 30\, \hbox{TeV}$. For `generic' models, $m_{\phi} \sim m_{3/2} \sim M_{soft}$, leading to a
tension with supersymmetric solutions of the hierarchy problem.

There is a cognate problem with respect to decays to gravitini \cite{ph0602061}. Even if $m_{\phi} \gg M_{soft}$, provided $m_{\phi} \gtrsim 2 m_{3/2}$ the decay mode $\phi \to \psi_{3/2} \psi_{3/2}$ is kinematically open.
This decay mode is problematic as for $m_{3/2} \lesssim 30 \,\hbox{TeV}$ the gravitino decays could affect the successful BBN predictions.

The second problem is to ensure that only the SM is reheated.
String theory generally contains many extra sectors in addition to the SM.
These include additional hidden gauge and matter sectors, as well as light axion-like particles.
Excessive branching ratios to these hidden sectors would lead to an overproduction of dark matter or $\Delta N_{eff} \gg 1$ and a failure of the BBN predictions.
There is also a practical difficulty. Calabi-Yaus have many - easily $\mc{O}(100)$ - moduli which in
generic models of moduli stabilisation have parametrically similar masses and lifetimes.
A study of reheating then requires a coupled analysis of all moduli and their decay modes.
Such an analysis is not only impractical, it is also highly sensitive to the post-inflationary initial conditions as the relative energy densities
in each modulus field depends on the magnitude of the initial modulus misalignment and its non-perturbative production rate at pre-heating.

\section{Dark Radiation in LVS Vacua}

We shall analyse reheating within the LARGE Volume Scenario (LVS) \cite{0502058}, in which the volume is stabilised at
exponentially large values.
For definiteness, we take a canonical strong Swiss-cheese form for the volume
\be
\mc{V} = \frac{1}{\lambda} \left( \tau_b^{3/2} - \sum_i \lambda_i \tau_{s,i}^{3/2} \right)\,,
\ee
but our conclusions would hold also for more general LVS scenarios.
For recent explicit constructions of LVS models see \cite{swisscheese}.

LVS has some particular advantages rendering this analysis meaningful.
First, the moduli masses take a distinctive hierarchy, with the volume modulus by far the lightest of the non-axionic fields ($a$ are axionic partners)
\bea
M_P & \equiv & 2.4 \ti 10^{18} \,\hbox{GeV}, \nn \\
M_{string} & \sim & M_P/\sqrt{\mc{V}}, \nn \\
m_{\tau_{s,i}} \sim m_{a_{s,i}} & \sim & M_P \ln \mc{V}/\mc{V}, \nn \\
m_{3/2} \sim m_{U} \sim m_{S} & \sim & M_P/\mc{V}, \nn \\
m_{\tau_b} & \sim & M_P/\mc{V}^{3/2}, \nn \\
m_{a_b} & \lesssim & M_P \,e^{-2 \pi \mc{V}^{2/3}} \sim 0\,. \nn
\eea
Of particular importance for this paper is the volume axion $a_b$. As it can only obtain mass by effects non-perturbative in the volume,
it is effectively massless on all cosmological scales. Its presence and masslessness are universal implications of LVS \cite{0502058, CGR}, and so
it is important to extract physical implications of $a_b$.

Second, as $\mc{V} \gg 1$ and $\tau \propto m^{-3}$, the volume modulus $\tau_b$ outlives all other moduli by factors polynomial in volume.
With radiation redshifting as $a^{-4}$ and matter as $a^{-3}$,
this implies any radiation produced by the decays of other moduli will be highly diluted
by the time the volume modulus decays.
In LVS it is therefore reasonable to expect that the
universe goes through a period when its energy density is dominated by coherent oscillations of the
volume modulus $\tau_b$, with reheating driven by the decays of $\tau_b$.

What is $m_{\tau_b}$?
Naively, in gravity mediation $M_{soft} \sim m_{3/2}$. However the no-scale susy breaking structure of LVS leads to large cancellations in the
soft terms, both at tree level and loop level (anomaly mediation \cite{cancelations, 09063297}, see however \cite{shanta}) \footnote{We shall not consider here
field redefinitions such as in \cite{10030388}. Independent of the subtle questions on the conditions required for their existence,
if present and with the effects estimated in \cite{10030388} they bring soft terms to the scale of the gravitino mass, restoring the full-blown CMP.}.
 In the case where the SM is realised by branes at a singularity \cite{09063297}, gaugino masses are suppressed to a scale
$M_{1/2} \sim M_P / \mc{V}^2$. For soft scalar masses,
the level of calculation in \cite{09063297} could only establish suppression of soft scalar
masses to a scale $m_0^2 \sim M_P^2/\mc{V}^3$. Going beyond this requires $\alpha'$ corrections to the K\"ahler potential that have not been computed (in particular, the $\zeta(3) \mc{R}^4$ correction in the \emph{matter} K\"ahler metric).
We shall assume the level of volume-sequestering is the same for both gaugino and scalar masses, and
so (for an analysis of the non-sequestered case see \cite{12072771})
$$
M_{soft} \sim \frac{M_P}{\mc{V}^2} \ll m_{\tau_b} \sim \frac{M_P}{\mc{V}^{3/2}}\,.
$$
Setting $M_{soft} \sim 1 \,\hbox{TeV}$ then leads to $m_{\tau_b} \sim 3 \ti 10^6 \,\hbox{GeV}$, comfortably heavy enough to avoid
the CMP.

\subsection{Reheating via volume modulus decays}

To study reheating in LVS we therefore focus on perturbative decays of the volume modulus
 (ref. \cite{StringReheating} focused instead on decays of the
inflaton).
We require the relative fractions of radiation entering hidden and SM degrees of freedom,
$f_{hidden}$ and $f_{SM} \equiv 1 - f_{hidden}$.
In turn this requires the couplings of the volume modulus - the overall breathing mode - to both visible and hidden degrees of freedom
(some moduli-matter couplings have already been computed in \cite{LVSatFiniteT}).

\subsubsection{Decays to volume axions}

The basic theory we start with is
\be
\label{pig}
K = - 3 \ln (T_b + \bar{T}_b)\,.
\ee
Although LVS is a complicated many-modulus model, eq. (\ref{pig}) captures the couplings of the volume modulus:
the small cycle moduli are much
smaller and much heavier than the overall volume modulus and any mixing is volume-suppressed. This gives
\be
\mc{L} = \frac{3}{4 \tau_b^2} \,\partial_{\mu} \tau_b \partial^{\mu} \tau_b + \frac{3}{4 \tau_b^2} \,\partial_{\mu} a_b \partial^{\mu} a_b\,.
\ee
We canonically normalise the volume modulus by writing $\Phi = \sqrt{\frac{3}{2}} \ln \tau_b$,
giving
\be
\label{goat}
\mc{L} =  \half \partial_{\mu} \Phi \partial^{\mu} \Phi + \half \left( \frac{3}{2} \exp[-2 \sqrt{\frac{2}{3}} \Phi] \right) \partial_{\mu} a_b \partial^{\mu} a_b.
\ee
Eq. (\ref{goat}) contains the self-coupling of the volume modulus to its corresponding axion. From (\ref{goat}) we can directly calculate
the decay rate $\Phi \to a_b a_b$, obtaining
\be
\label{axiondecay}
\Gamma_{\Phi \to a_b a_b} = \frac{1}{48 \pi} \frac{m_{\Phi}^3}{M_P^2}.
\ee
As the bulk axion only interacts via Planck-suppressed couplings \cite{CGR}, the decays to $a_b$ correspond to dark radiation.

\subsubsection{Decays to visible sector fields}

Let us consider decays to MSSM particles.
Here we are using the expression `MSSM' but the exact model is not so crucial and the considerations apply more generally.

\emph{Gauge bosons}: Couplings of the volume modulus to gauge bosons arise through the dependence of the gauge kinetic function on $T_b$.
However as the MSSM is realised locally, the tree-level gauge couplings $f_a = S + h_{a,i} T_{s,i}$ are independent of $T_b$.
Such a dependence is certainly induced radiatively, as the volume determines the string scale and hence the high scale from which couplings
start running. At loop level there is then a term $\frac{b_a \alpha_{SM}}{4 \pi} \ln \mc{V}$ in the gauge kinetic function, which induces a coupling
\be
\frac{\lambda_a \alpha_{SM}}{4 \pi} \,\Phi F_{\mu \nu} F^{\mu \nu} + \ldots
\label{DecayGauge}
\ee
However as a radiative correction this only gives $\Gamma_{\Phi \to A^{\mu}_{SM} A^{\mu}_{SM}}  \sim \left( \frac{\alpha_{SM}}{4 \pi} \right)^2 \frac{m_{\Phi}^3}{M_P^2}$,
which is highly subdominant compared to eq. (\ref{axiondecay}).

\emph{Matter scalars}: The couplings to MSSM matter scalars $C$ are set by the K\"ahler potential, which extends (\ref{pig}) to
\be
\label{pig2}
K = - 3 \ln (T_b + \bar{T}_b) + \frac{C \bar{C}}{(T_b + \bar{T}_b)}\,.
\ee
The $\left(T_b + \bar{T}_b\right)^{-1}$ dependence is fixed by the local nature of the MSSM.
After canonical normalisation the relevant modulus-matter interaction is
\be
\label{pig3}
\frac{1}{2} \sqrt{\frac{2}{3}}\, \Phi \left( \bar{C} \Box C + C \Box \bar{C} \right) \,.
\ee
Note that to eliminate $\Phi \partial_{\mu} C \partial^{\mu} C$ couplings it is crucial to use the $K_{C\bar{T}}$ terms.
This interaction also vanishes for on-shell massless particles, giving $\Gamma_{\Phi \to C\bar{C}} \sim \frac{m_0^2 m_{\Phi}}{M_P^2} \ll \frac{m_{\Phi}^3}{M_P^2}$.

\emph{Matter fermions, gauginos and Higgsinos}: The couplings to fermions can also be derived from the K\"ahler potential (\ref{pig2}) and turn out to be of the form
\be
\label{sheep}
\lambda \frac{\Phi}{M_P} \,\bar{\chi} \bar{\sigma}^m D_m \chi\,.
\ee
However such couplings are chirality suppressed and give $\Gamma_{\Phi \to ff} \sim \frac{m_f^2 m_{\Phi}}{M_P^2} \ll \frac{m_{\Phi}^3}{M_P^2}$.
Actually, as for gauge bosons we expect this decay mode to be generated radiatively,
again giving a subdominant contribution
$\Gamma \sim \left( \frac{\alpha_{SM}}{4 \pi} \right)^2 \frac{m_{\Phi}^3}{M_P^2}$.

\emph{Higgs bosons}: We finally consider decays to Higgs bosons. These, uniquely, have the possibility of a Giudice-Masiero coupling in the K\"ahler potential
\be
K = - 3 \ln (T_b + \bar{T}_b) + \frac{H_u \bar{H}_u  + H_d \bar{H}_d}{(T_b + \bar{T}_b)}
+ \left( \frac{Z H_u H_d}{(T_b + \bar{T}_b)} + \hbox{h.c.} \right). \nn
\ee
Here $Z$ is an undetermined constant (to leading order in an inverse volume expansion).
If the Higgs sector has a shift symmetry (as considered in \cite{12042551}), then $Z=1$ since
\be
K = \ldots + \frac{(H_u + \bar{H}_d)(\bar{H}_u + H_d)}{(T_b + \bar{T}_b)} + \ldots\,. \nn
\ee
At string scale energies the Giudice-Masiero term can be forbidden by anomalous $U(1)$s, in which case this term is generated
 after breaking to the SM. After canonical normalisation the resulting Lagrangian is
\bea
\label{Higgsdecays}
\mc{L} & = & \half \,\partial_{\mu} \Phi \partial^{\mu} \Phi + \partial_{\mu} H_u \partial^{\mu} \bar{H}_u
+ \partial_{\mu} H_d \partial^{\mu} \bar{H}_d  \\
&+& \frac{1}{\sqrt{6}} \left[\,\Phi\left( \bar{H}_u  \Box H_u + \bar{H}_d \Box H_d  \right) + Z H_u H_d  \Box \Phi + h.c.\right]. \nn
\eea
The last term in the second line of eq. (\ref{Higgsdecays}) allows for the direct decay $\Phi \to H_u H_d$, with
\be
\label{gammahiggs}
\Gamma_{\Phi \to H_u H_d} = \frac{2 Z^2}{48 \pi} \frac{m_{\Phi}^3}{M_P^2}\,.
\ee
The decay $\Phi \to H_u H_d$ is the only unsuppressed, and therefore dominant, MSSM decay mode.
There will also be subleading contributions to eq. (\ref{gammahiggs}) from SM radiative corrections and running of the interaction
strength from $M_{string}$ to $m_{\Phi}$.

\subsubsection{Decays to other axions}

\emph{Local closed string axions}: Another source of light fields are closed string axions, the imaginary parts of K\"ahler moduli.
The existence of such axions is model dependent as they
can be dynamically lifted by non-perturbative effects. Take the K\"ahler potential
\be
K = - 3 \ln (T_b + \bar{T}_b) + \frac{Z_{ax}}{2} \frac{(X + \bar{X})^\gamma}{(T_b+ \bar{T}_b)^{\lambda}}.
\ee
Here the axion is $\hbox{Im}(X)$ and for a local closed string axion $\lambda = 3/2$. Moreover,
$\gamma =2$ if the saxion $\hbox{Re}(X)$ is shrunk at a singularity while $\gamma =3/2$ if the saxion is in the geometric regime.
Writing $X = \frac{1}{\sqrt{2}}(A + i B)$, after normalisation the relevant couplings are
\bea
\mc{L} & = & \half \partial_{\mu} \Phi \partial^{\mu} \Phi + \half \partial_{\mu} A \partial^{\mu} A + \half \partial_{\mu} B \partial^{\mu} B \nn \\
& & + \sqrt{\frac{2}{3}} \frac{\lambda}{4 M_P} \left( A^2 - B^2 \right) \Box \Phi\,.
\eea
Note $Z_{ax}$ has disappeared from the Lagrangian and for $\gamma=3/2$ one would have additional
cross couplings $\Phi \partial_{\mu} a \partial^{\mu} B$ which are suppressed by $\mc{V}^{-7/6}$,
and so can be neglected. This gives rise to
\be
\Gamma_{\Phi \to BB} = \left( \frac{\lambda}{3/2} \right)^2 \frac{9}{16} \frac{1}{48 \pi} \frac{m_{\Phi}^3}{M_P^2}\,,
\ee
which if present is an extra and competitive source of hidden radiation.

\emph{Open string axions}: Note that the presence of a QCD axion does not guarantee
the existence of local closed string axions, as the QCD axion could arise from the angular component $\theta$ of open string matter fields
$C =\rho\,e^{i\, \theta}$ which obtain a vev $\langle\rho\rangle\neq 0$.
However, decays to open string axions $\theta$ can be shown to be suppressed
as they would be induced by terms of the form
\be
- \sqrt{\frac{2}{3}} \left(\frac{\langle \rho \rangle}{M_P}\right)^2 \Phi \theta \Box \theta
 + \frac{4}{3} \left(\frac{\langle \rho \rangle}{M_P}\right)^2 \Phi \partial_{\mu} a_b \partial^{\mu} \theta\,.
\label{openaxions}
\ee
The first term in (\ref{openaxions}) gives rise to $\Phi \to \theta \theta$ decays which are mass suppressed
while the second term yields $\Phi \to \theta a_b$ decays which could compete
with the $\Phi \to a_b a_b$ decays for $\langle \rho \rangle \sim M_P$.
However in general $\rho$ is fixed by D-terms and its vev is set by the Fayet-Iliopoulos term $\xi$
\be
(\langle\rho\rangle/M_P)^2 \sim \xi \sim 1/\mc{V} \ll 1\,.
\ee
Axions as dark radiation have been considered in a different context in \cite{axiondr}.

\subsubsection{Decays to hidden sector fields}

\emph{Gauge bosons on the large cycle}: Another source of competitive decays are from light gauge bosons from branes wrapping the bulk cycle.
Such branes might be present if an O7-plane wraps this cycle.
In this case
the bulk modulus has the direct coupling
\be
\tau_b F_{\mu \nu} F^{\mu \nu} \to \frac{1}{4} F_{\mu \nu} F^{\mu \nu} + \frac{1}{4} \sqrt{\frac{2}{3}} \frac{\Phi}{M_P} F_{\mu \nu} F^{\mu \nu},
\ee
which induces a decay rate
\be
\Gamma_{\Phi \to A^{\mu}_{bulk} A^{\mu}_{bulk}} = \frac{n_g}{2} \frac{m_{\phi}^3}{48 \pi M_P^2},
\ee
where $n_g$ is the number of gauge generators.
Such a bulk gauge group would represent a new hyperweak interaction, with $\alpha_{bulk} \lesssim 10^{-4}$.
Note that decays to Higgses and superpartners would be kinematically forbidden since these sectors would not be
sequestered, and so $M_{soft}^{hidden}\sim m_{3/2}\gg m_{\Phi}$.

\emph{Gauge bosons on small cycles}: In this case the decays to gauge bosons and gauginos
would be kinematically forbidden in the presence of strong dynamics at a high scale (gaugino condensation).
Hence the only worrisome decays are to light gauge bosons but they
would be loop-suppressed similar to eq. (\ref{DecayGauge}).

\emph{Bulk closed string $U(1)$s}: One model-independent source of massless gauge bosons are the bulk closed string $U(1)$s that arise from reduction of $C_4$ along 3-cycles. In general there are $h^{1,2}_+$ such $U(1)$s, where $h^{1,2}_+$ is the number of 3-cycle invariant under the orientifold projection. However the gauge kinetic function of these is set by the complex structure moduli, giving
\be
\label{cow}
\mc{L} = f(U) F_{\mu \nu} F^{\mu \nu} + f(U) \lambda \sigma^{\mu} \mc{D}_{\mu} \bar{\lambda}.
\ee
As mixing between the K\"ahler and complex structure moduli is volume-suppressed at large volumes, $\tau_b$ only couples to such $U(1)$s
via volume-suppressed mixing terms, and so such decay modes are negligible.

\emph{Sequestered hidden sectors}: Other competing decay rates are to Higgs bosons belonging to sequestered hidden sectors at different singularities
since they would behave as eq. (\ref{gammahiggs}).

\subsubsection{Summary of leading decay channels}

Let us summarise all the leading decay channels for the volume mode:
\begin{enumerate}
\item Volume axion
\item Visible sector Higgses
\item Local closed string axions (if present)
\item Light gauge bosons on the large cycle (if present)
\item Higgses in sequestered hidden sectors (if present)
\end{enumerate}

\subsection{Predictions for dark radiation}

Given that the presence of the volume axion and visible Higgses is model-independent,
let us start focusing only on the decays $\Phi \to a_b a_b$ and $\Phi \to H_u H_d$. The relative fraction of hidden radiation is
\be
f_{hidden} = \frac{1}{1 + 2Z^2}\,.
\ee
Let us now relate $f_{hidden}$ to $\Delta N_{eff}$. Conservation of comoving entropy $s = g(T) a^3 T^3$ implies
the SM temperature $T_{SM}$ and energy density $g(T) T^4$ behave as
$$
T_{SM} \sim \frac{1}{g^{1/3} a}\,, \qquad \rho_{SM} = \frac{\rho_{SM}^{init}}{g^{1/3} a^4}\,.
$$
The ratio of hidden to SM radiation at neutrino decoupling temperature $T_{dec}$ is then
$$
\frac{\rho_{hidden}}{\rho_{SM}} = \frac{\rho_{hidden}^{init}}{\rho_{SM}^{init}} \left( \frac{g(T_{dec})}{g(T_{reheat})} \right)^{1/3}.
$$
The number of extra effective neutrino species is
\bea
\Delta N_{eff} & = & 3 \frac{\rho_{hidden}}{\rho_{neutrinos}} =  \frac{43}{7} \frac{\rho_{hidden}}{\rho_{SM}} \nn \\
&  = & \frac{43}{7} \frac{f_{hidden}}{1-f_{hidden}} \left( \frac{g(T_{dec})}{g(T_{reheat})} \right)^{1/3}.
\eea
The reheating temperature can be easily estimated as
\be
T_{reheat} \sim \sqrt{\Gamma_\Phi M_P} \sim \mc{O}(1) \,\,\hbox{GeV} \left(\frac{m_{\Phi}}{3 \ti 10^6 \,\hbox{GeV}}\right)^{3/2}\,. \nn
\ee
For $T_{reheat}\lesssim 1$ GeV, $g(T_{reheat}) = 247/4$, whereas for $T_{reheat}\gtrsim 5$ GeV, $g(T_{reheat}) = 345/4$,
generating a small uncertainty in the final prediction for $\Delta N_{eff}$ which for $g(T_{dec}) = 10.75$ is
\be
3.12\, \kappa \leq  \Delta N_{eff}   \leq  3.48 \,\kappa
\label{prediction}
\ee
where $\kappa = f_{hidden}/(1 - f_{hidden})= 1/(2 Z^2)$.
For $Z=1$, $\kappa = 1/2$, and so we obtain $1.56 \leq \Delta N_{eff} \leq 1.74$ which is comparable in magnitude to the observational hints.

\begin{figure}
\begin{center}
\includegraphics[width=8cm]{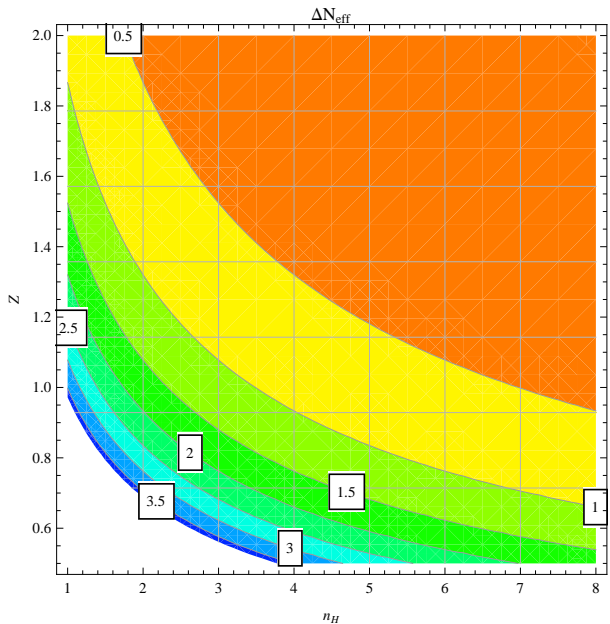}
\end{center}
\vspace{-5ex}
\caption{Contour plot of $\Delta N_{eff}$ versus $n_H$ ($x$-axis) and $Z$ ($y$-axis) for $n_a=0$
and $T_{reheat}\lesssim 1$ GeV.\vspace{-3ex}}
\label{Figure}
\end{figure}

If we allow a generic number of Higgs doublets $n_H$ and local closed string axions $n_a$,
the expression for $\kappa$ generalises to $\kappa=(1+9 n_a/16)/{n_H Z^2}$. In figure \ref{Figure}
we present a plot of $\Delta N_{eff}$ versus $n_H$ and $Z$ for $n_a=0$ (and so the QCD axion has to be an open string mode).
If $n_a \neq 0$, and in particular $n_a \sim \mc{O}(100)$ in the case of an `axiverse' \cite{LVSaxions}, $\Delta N_{eff}$
quickly grows above the allowed experimental bounds.

Let us briefly consider other possible sources of hidden radiation.
If light gauge bosons on $\tau_b$ exist, there are various
possibilities:
\begin{itemize}
\item If the bulk gauge group is not in thermal equilibrium with the SM and is not Higgsed, its gauge bosons constitute dark radiation.

\item If the bulk gauge group is not in thermal equilibrium with the SM and is Higgsed, its massive gauge bosons can
constitute dark matter if they are stable.

\item If the bulk gauge group is in thermal equilibrium with the SM
(which might be obtained via kinetic mixing of bulk $U(1)$s with the ordinary photon),
then the bounds on $\Delta N_{eff}$ require it to be Higgsed at a scale
$\gtrsim 1 \,\hbox{MeV}$, and this mode counts as SM radiation.
\end{itemize}

Notice that also hidden photons produced by photon $\leftrightarrow$
hidden photon oscillations in the thermal bath could form dark radiation
\cite{Andreas}.
Finally, in the presence of hidden sequestered scenarios at different singularities,
it is very likely that $\Delta N_{eff}$ would become too large and dark matter would be
overproduced.

\section{Conclusions}

The main point to emphasise is that dark radiation is generic and unavoidable in the sequestered LARGE Volume Scenario. This relies only on reheating being driven by decays of the lightest modulus, which always has an open decay mode to its axion partner. The magnitude of this dark radiation depends on assumptions about the visible sector and the number of additional closed string axions, but can easily be at a level consistent with observational hints for $\Delta N_{eff}$.
The bounds on $\Delta N_{eff}$ can be used to strongly constrain the couplings and matter content of the models,
showing that it is very hard to achieve an axiverse in sequestered models. Future observations, such as those expected from PLANCK,
will further constrain this scenario.

\emph{Note added: This paper is submitted simultaneously to the related work \cite{TakahashiHigaki}.}
\subsection*{Acknowledgments}

JC is supported by the Royal Society with a University Research Fellowship. We thank Jo Dunkley, Mark Goodsell, Sven Krippendorf, Anshuman Maharana, Andreas Ringwald, Timo Weigand and particularly Tetsutaro Higaki and Fuminobu Takahashi for discussions.
JC thanks the organisers of the 3rd UT Quest Workshop on Superstring Cosmophysics in Hokkaido for a very stimulating atmosphere where crucial progress was made.
We also thank the Isaac Newton Institute program for hospitality during the BSM program.

\bibliography{apssamp}

\end{document}